\documentclass[twocolumn,preprintnumbers,amsmath,amssymb,aps,nobibnotes,nofootinbib]{revtex4} %superscriptaddress
\pdfoutput=1

\usepackage{graphicx}
\graphicspath{{Figs/}}
\usepackage{dcolumn}% Align table columns on decimal point
\usepackage{bm}% bold math
\usepackage[a4paper, top=1.0in, bottom=1.2in, left=1.0in, right=1.0in]{geometry}
\usepackage{multirow}

\newcommand{\<}{\langle}
\renewcommand{\>}{\rangle}
\newcommand{\beq}{\beqn}
\newcommand{\eeq}{\eeqn}
\newcommand{\beqn}{\begin{eqnarray}}
\newcommand{\eeqn}{\end{eqnarray}}
\def\sla#1{\setbox0=\hbox{$#1$}\dimen0=\wd0
      \setbox1=\hbox{/} \dimen1=\wd1 \ifdim\dimen0>\dimen1
      \rlap{\hbox to \dimen0{\hfil/\hfil}} #1                        \else
      \rlap{\hbox to \dimen1{\hfil$#1$\hfil}}
      /   \fi}

\newcommand{\nn}{\nonumber}

\newcommand{\mc}{\mathcal}
\newcommand{\glr}{$G_{{\rm LR}}$}
\newcommand{\id}{1 \hspace{1.15mm} \!\!\!\!1}
\newcommand{\ka}{\kappa}
\newcommand{\kp}{\kappa'}

\newcommand{\tr}{{\rm Tr}}
\newcommand{\hc}{{\rm h.c.}}

\long\def\symbolfootnote[#1]#2{\begingroup%
 \def\thefootnote{\fnsymbol{footnote}}\footnote[#1]{#2}\endgroup}

\newcommand{\noi}{\noindent}

\begin{document}

\preprint{TUM-HEP-768/10}\preprint{UMD-PP-10-012}

\title{\boldmath TeV Scale Left-Right Symmetry and Flavor Changing Neutral Higgs Effects}

\author{Diego~Guadagnoli}
\affiliation{Excellence Cluster Universe, Technische Universit\"at M\"unchen, Boltzmannstra{\ss}e 2, D-85748 Garching, Germany}

\author{Rabindra~N.~Mohapatra}
\affiliation{Maryland Center for Fundamental Physics, Department of Physics, University of Maryland, College Park, MD 20742, USA}

\date{\today}

\begin{abstract}
\noindent In minimal left-right symmetric models, the mass of
the neutral Higgs field mediating tree-level flavor changing
effects (FCNH) is directly related to the parity breaking scale.
Specifically, the lower bound on the Higgs mass coming from 
Higgs-induced tree-level effects, and exceeding about 15 TeV, 
would tend to imply a $W_R$ mass bound much higher than that 
required by gauge exchange loop effects -- the latter allowing 
$W_R$ masses as low as 2.5 TeV.
Since a $W_R$ mass below 4 TeV is accessible at the LHC, it is
important to investigate ways to decouple the FCNH effects from 
the $W_R$ mass. In this Letter, we present a model where this happens, 
providing new motivation for LHC searches for $W_R$ in the $1-4$ TeV 
mass range.
\end{abstract}

\maketitle

\section{Introduction}

\noi Left-right symmetric models (LRSM), initially proposed to
explain the origin of parity violation \cite{LR}, have received
further motivation from the fact that they provide a natural
setting for understanding small neutrino masses via the seesaw
mechanism \cite{seesaw}. The seesaw scale in these models happens
to be the scale of parity breaking, thereby connecting the
smallness of neutrino masses to the dominance of V$-$A
interactions in low energy weak interactions. This connects two
independently motivated new physics scales beyond the Standard
Model (SM). Another compelling reason for considering this class of
models is the Gell-Mann--Nishijima-like formula that relates
electric charge, weak isospins and the baryon and lepton numbers
of particles \cite{marshakmohapatra} as follows:
\beq
Q = T_{3L} + T_{3R} + (B-L)/2~,
\eeq
thereby justifying the hypercharge quantum numbers -- that in
the SM have just ad-hoc values -- as remnants
of the $SU(2)_R$ and $B-L$ gauge symmetries.

A question of great current interest is whether the new gauge
bosons associated with the LRSM can be observed at the CERN LHC.
Since the $W_R$ has an unmistakable signature at the LHC (of two
like sign dileptons and two jets, with no missing energy)
\cite{keung}, and a large enough production cross section if its
mass is below 4 TeV \cite{gninenko}, one needs to know any lower
bounds on its mass from measured low energy observables.  The
relevant processes are the SM suppressed ones, in particular
meson-antimeson mixing and CP violation in the kaon sector.
Extensive analyses of these
constraints have been carried out \cite{cons} over the years and
the latest results can be stated as follows: the most stringent
bounds come from the CP violating observables $\epsilon$ and
$\epsilon'$ \cite{cons}, and possibly the neutron electric dipole
moment \cite{ji}; they however depend on how CP violation is
incorporated into the model. For example, in the minimal
versions of the model the left and right CKM angles are nearly
the same, and the bounds from CP violating observables can
even be absent altogether, if one defines parity as transforming 
$Q_L\to Q^c_L$ \cite{goran}. The next strongest bounds come, 
primarily, from the $K_L - K_S$ mass difference, and arise from 
new, gauge mediated
contributions. Specifically, defining parity as $Q_L\to Q_R$, 
one finds a bound of $4$ TeV \cite{Zhang}, whereas with the 
$Q_L\to Q^c_L$ parity definition, one gets $2.5$ TeV \cite{goran}.
The weaker of the above bounds still allows the LHC to search for 
the $W_R$ and the associated $Z'$.

There is however another source of flavor violation in left-right
models \cite{fcnh}. Even in its minimal version, a LRSM contains two
copies of the standard model Higgs doublet, embedded into the
bi-doublet needed to generate fermion masses. This is
therefore a version of two Higgs models which can be
characterized, in the notation of \cite{carlucci}, as having
$Y_{2d}= Y_{1u}$ and $Y_{2u}=Y_{1d}$. As such, these models give
rise to tree-level flavor changing neutral Higgs (FCNH) effects
coming from the neutral member of the second SM Higgs doublet
contained in the bi-doublet. We will call this the FCNH problem of
LRSM. These effects turn out to put a lower limit on the second
Higgs mass of around $10 - 15$ TeV. This bound would imply a
corresponding bound on the parity breaking scale. In fact,
the Higgs sector of minimal left-right models has been analyzed
extensively \cite{boris} and it
has been shown that, with the most general Higgs potential, the
masses of the Higgs fields belonging to the second doublet owe
their origin to the parity breaking scale $M_{W_R}/g$.
Furthermore, if the Higgs self scalar couplings are kept in the
perturbative range (say $\lambda_i \leq 1$), the FCNH constraints
would raise the right handed $W_R$ scale to the 10 TeV ballpark,
pushing it out of the LHC reach.

So the immediate question that arises is whether it is possible,
and how naturally, to satisfy all the FCNH constraints while
keeping the parity breaking scale accessible at the LHC. Some
solutions to this problem have been suggested in the literature
\cite{other} by invoking supersymmetry and cancelling the above
effects by supersymmetric box contributions or by using special CP
violating solutions. Another route, where manifest
left-right symmetry is forsaken, is to have the right handed
quark-mixing angles different from the left handed ones
\cite{buras}. The latter possibility (the so-called non-manifest
LRSM), which has also been used to lower the bounds on the $W_R$
mass itself \cite{non}, requires an extended Higgs sector.

In this Letter, we seek an alternative solution to the FCNH problem
of LRSM within the non-supersymmetric framework via a minimal
extension of the Higgs sector, without sacrificing the manifest
nature of left and right handed quark mixings. Within our solution,
the contributions to quark masses that break the relation 
${\rm diag}(m_u,m_c,m_t)$ $\propto$ ${\rm diag}(m_d,m_s,m_b)$,
as well as flavor mixing ($V^{\rm CKM} \neq \id$), both originate 
from an effective operator of dimension 5.
It is straightforward to extend our model to the supersymmetric case.

The philosophy behind our approach is that the Higgs sector is
independent of the gauge sector of generic models and their
effects should in principle be controlled by separate physics. In
particular, in the context of the left-right models, the Higgs
sector should not be the determining factor as far as the scale of
parity violation is concerned. Thus the FCNH problem of minimal
LRSM may be a low energy manifestation of the fact that the model
is incomplete and needs extension. Turning this question around,
if indeed a low, TeV scale, $W_R$ is discovered at the LHC, it
would be an indication that the minimal LRSM needs extension in a
way such that gauge and Higgs sectors are separated. We show in
this Letter that it is indeed possible to separate the Higgs sector
constraints from those of the gauge sector by introducing a
discrete symmetry and a second bi-doublet with $B-L=2$ with mass
$M_\rho$ much higher than the left-right scale. The FCNH effects
of the minimal left-right model are shown to be associated with
this new mass scale, thus allowing the parity breaking scale to be
determined solely by the $W_R$ exchange effects at the one loop
level. Although our goal is not to construct a grand unified
theory, we note, incidentally, that the new Higgs bi-doublet in
our model is part of the SO(10) multiplet {\bf 210} often used to
break GUT symmetries. Below $M_\rho$ and the scale of parity breaking,
this model resembles a two Higgs doublet model such as the
Glashow-Weinberg model \cite{GW} and therefore has natural flavor
conservation. The suppression of FCNH effects can now be explained
keeping all scalar couplings perturbative and without
simultaneously ``dragging'' the $W_R$ and $Z'$ mass scale to the
$M_\rho$ bound.

This Letter is organized as follows: in sec. \ref{sec:LRSMmin}, we review
the salient features of the minimal left-right symmetric model and the
associated FCNH problem; in secs. \ref{sec:LRSMrho} and \ref{sec:quarkmix},
we present the new extended model and the mechanism for generating
quark masses and mixings; in sec. \ref{sec:FCNH}, we discuss the associated
FCNH effects and demonstrate the decoupling of the FCNH scale from
the $W_R$ scale. In sec. \ref{sec:comments}, we make some further 
remarks on the model.

\section{\boldmath The Left-Right Symmetric Model: Motivation and
Minimal Realization} \label{sec:LRSMmin}

\noi Left-right-symmetric models \cite{LR} are based on the group
\glr $\equiv SU(2)_L \times SU(2)_R \times U(1)_{B-L}$, with
parity assumed to be a good symmetry of the Lagrangian. Let us
first summarize their minimal realization. The SM matter
content can be intuitively accommodated into fundamental
representations of the $G_{LR}$ group, namely
\beq
\label{eq:fermions}
\begin{tabular}{c c}
  \hline
{\bf field} & {\bf ~~~$\mathbf{G_{{LR}}}$ representation~~~} \\
  \hline
 & \\
[-0.3cm]
  $Q_L$ = $\left( \begin{array}{c}
                u_L \\ d_L
               \end{array} \right)$ & $(2,~1,~1/3)$ \\
[0.4cm]
  $Q_R$ = $\left( \begin{array}{c}
                u_R \\ d_R
               \end{array} \right)$ & $(1,~2,~1/3)$ \\
[0.4cm]
  $L_L$ = $\left( \begin{array}{c}
                \nu_L \\ \ell_L
               \end{array} \right)$ & $(2,~1,~-1)$ \\
[0.4cm]
  $L_R$ = $\left( \begin{array}{c}
                \nu_R \\ \ell_R
               \end{array} \right)$ & $(1,~2,~-1)$ \\
[0.25cm]
  \hline
\end{tabular}
\eeq
In particular, right handed neutrinos are introduced naturally
by the requirement that right handed particles be also in doublets.
These models therefore offer a very natural set-up for realizing
the seesaw mechanism for small neutrino masses. The gauge currents
associated with $G_{LR}$ couple with strengths $g_L, g_R$ and $g'$,
with $g_L = g_R$ to guarantee parity symmetry at higher energies.

The $G_{LR}$ gauge group is broken spontaneously by the vev's
of an appropriate Higgs sector. The Higgs sector most suitable for
implementing the seesaw mechanism is given by a bi-doublet
$\phi \sim (2,~2,~0)$ and two triplets $\Delta_L \sim (3,~1,~2)$
and $\Delta_R \sim (1,~3,~2)$, namely
\beq
\label{eq:phiDelta}
&&\phi = \left( \begin{array}{cc}
                \phi_1 & \phi_2 \\
               \end{array} \right)
       = \left( \begin{array}{cc}
                \phi_1^0 & \phi_2^+ \\
                \phi_1^- & \phi_2^0 \\
               \end{array} \right)~, \nn \\
&&\Delta_{L(R)} = \left( \begin{array}{cc}
                \delta_{L(R)}^+/\sqrt 2 & \delta_{L(R)}^{++} \\
                \delta_{L(R)}^0 & -\delta_{L(R)}^+/\sqrt 2 \\
               \end{array} \right)~.
\eeq
$G_{LR}$ is broken first to $SU(2)_L \times U(1)_Y$ by the
$\< \Delta_R \>$ vev: we assume this scale to lie in the few TeV range.
Electroweak symmetry breaking is then induced by the vev's of
$\phi$.

In left-right models, there are two different ways to define the
parity operation: {\em (i)} $Q_L\leftrightarrow Q_R$, {\em (ii)}
$Q_L\leftrightarrow Q^c_L$. The second definition arises naturally
in the context of SO(10) grand unified models as well as
supersymmetric left-right models. We will illustrate our new model
using the first definition of parity although the results can be
extended in a straightforward manner to the second case.

In the realization of parity invariance where $Q_L\leftrightarrow Q_R$,
the four vev's associated with the $\phi$ and $\Delta_{L,R}$ fields can
be complex: one can however perform \cite{Zhang} two field
redefinitions, thereby setting two of the four phases to zero.
Hence, in all generality, the Higgs vev's can be chosen as
\beq
\label{eq:vevs}
&&\<\phi\> = \frac{1}{\sqrt2}\left( \begin{array}{cc}
                \kappa & 0 \\
                0 & \kappa' e^{i\alpha} \\
               \end{array} \right)~, \nn \\
&&\<\Delta_L\> = \frac{1}{\sqrt2}\left( \begin{array}{cc}
                0 & 0 \\
                v_L e^{i \theta_L} & 0 \\
               \end{array} \right)~, \nn \\
&&\<\Delta_R\> = \frac{1}{\sqrt2}\left( \begin{array}{cc}
                0 & 0 \\
                v_R & 0 \\
               \end{array} \right)~.
\eeq
The quark fields and the Higgs bi-doublet in eqs. (\ref{eq:fermions})
and (\ref{eq:phiDelta}) give rise to the following Yukawa Lagrangian
\beq
\label{eq:LYgen}
\mc L_Y = \overline{\hat Q}_{Li} (h_{ij} \phi + \tilde h_{ij} \tilde \phi)
\hat Q_{Rj} + \hc
\eeq
where $\tilde \phi = -i \tau_2 \phi^* i \tau_2$, $i,j$ are flavor indices
and the hats indicate gauge eigenstates, to be distinguished from mass
eigenstates. Parity symmetry, under which $Q_L \to Q_R$ and
$\phi \to \phi^\dagger$, requires the Yukawa matrices $h$ and
$\tilde h$ to be Hermitian. Recalling that $\phi$ is a bi-doublet,
it is clear that the terms on the r.h.s. of eq. (\ref{eq:LYgen})
are all those present in the most general two Higgs doublet model
(2HDM), except for the fact that the four Yukawa-coupling matrices
allowed in the general case are here reduced to two, $h$ and
$\tilde h$, because of the left-right symmetry.

The Yukawa Lagrangian in eq. (\ref{eq:LYgen}), after spontaneous
symmetry breaking, gives rise to the quark mass matrices.
Specifically, after performing the field redefinitions
\beq
\label{eq:UDrotations}
\hat U_{L,R} = V^U_{L,R} U_{L,R}~,~~~\hat D_{L,R} = V^D_{L,R} D_{L,R}~,
\eeq
the diagonal mass matrices for up- and down-type quarks read
\beq
\label{eq:MUMD}
&&M_U = \frac{1}{\sqrt2} V^{U \dagger}_L \left( \kappa h
+ e^{- i \alpha} \kappa' \tilde h \right) V^U_R~,\nn \\
&&M_D = \frac{1}{\sqrt2} V^{D \dagger}_L
\left( e^{i \alpha} \kappa' h + \kappa \tilde h \right) V^D_R~.
\eeq
The above mass matrices must reproduce, e.g., the relation
$m_t \gg m_b$. Assuming no large cancellations between the
two terms on either r.h.s. of eqs. (\ref{eq:MUMD}), this implies
the hierarchies $\kappa \gg \kappa'$ and $h \gg \tilde h$,
namely that the first term in the expression for $M_U$ will
be dominant, $M_U \simeq \kappa h$, and one can always
choose a basis where this matrix is diagonal. In this basis,
the rotation on the $d$-quark fields that makes the $M_D$
matrix diagonal would be
\beq
\label{eq:MDrotation}
\hat D_{L,R} = V^{\rm CKM}_{L,R} D_{L,R}\,,~
\mbox{with }V^{\rm CKM}_{L,R} = V^{U \dagger}_{L,R} V^D_{L,R}\,.
\eeq

One can easily convince oneself that, after the above rotations
and after expressing the $h$ and $\tilde h$ couplings in terms of
the quark mass matrices, the $U$-$U$-Higgs ($D$-$D$-Higgs)
couplings in eq. (\ref{eq:LYgen}) that are proportional to the
$M_D$ ($M_U$) matrix, will be off-diagonal in flavor space because
of the CKM misalignment, and give rise (among the other effects)
to new tree-level contributions to flavor changing neutral current
processes. In addition, the presence of the new phases implies new
CP violating effects as well. In fact, a detailed analysis of
these effects has been performed in ref. \cite{Zhang}. Meson
mixings receive new loop contributions because of the $W_R$,
entailing a lower bound on $M_{W_R}$ around 4 TeV, or $2.5$ TeV
with an alternative parity definition. The already mentioned
quark-quark-Higgs couplings do also contribute to these processes
via tree-level diagrams mediated by the physical states $H_1^0$
and $A_1^0$. Assuming degeneracy in their masses, one obtains a
lower bound of about 15 TeV. These bounds get in general even more
severe once one takes into account also effects on CP violating
observables, primarily $\epsilon_K$. However, the CP violating
effects are more model-dependent, and we omit them here for the
sake of simplicity.

The above discussion is meant to highlight that, within the
minimal LRSM realized as above, the lowest allowed parity breaking
scale is typically determined by the FCNH mediated by the Higgs
fields rather than by the new gauge boson exchanges, thereby
pushing this scale out of the LHC range. In the next section we
will show how the inclusion of an additional assumption is able to
basically eliminate the above problems altogether, decoupling the
parity breaking scale from the FCNH scale and keeping the parity
breaking scale within reach of the LHC.

\section{\boldmath LRSM extension with a discrete symmetry} \label{sec:LRSMrho}

\subsection{\boldmath Definition}

\noi Looking at the Yukawa Lagrangian in eq. (\ref{eq:LYgen}), it is
evident that root of the severe FCNH problem inherent in the
general LRSM is the mentioned presence of four Yukawa couplings,
as they would appear in a general 2HDM \cite{haber}, but locked in
two pairs, $h$ and $\tilde h$, because of the LR symmetry. This
interlocking makes it impossible to implement the minimal flavor
violation hypothesis \cite{MFV} in the LRSM. In order to suppress
the FCNH effects, we therefore adopt the following procedure.

First, we note that the $\tilde h$ couplings would be forbidden in
presence of a discrete symmetry, e.g. $Z_4$ defined as
\beq
\label{eq:z4def}
&& \phi \stackrel{\rm Z_4}{\longrightarrow} i \phi~,
~~~ Q_R \stackrel{\rm Z_4}{\longrightarrow} -iQ_R~,\nn \\
&& L_L \stackrel{\rm Z_4}{\longrightarrow}-iL_L ~,
~~~ \Delta_L\stackrel{\rm Z_4}{\longrightarrow} -\Delta_L~,
\eeq
with all the other fields unchanged.\footnote{%
The possibility of the introduction of a horizontal symmetry, instead,
has been discussed In the context of general (i.e. not LR symmetric)
2HDMs and in the LRSM in \cite{BGL} and \cite{KAP} respectively.}

In this case, only the coupling matrix $h \neq 0$, whereas $\tilde
h~=~0$, implying $d$-quark mass terms of the form $M_D =
\kappa'/\kappa \, M_U$, which obviously cannot work for all the
$m_{d,s,b}$ masses. Besides, the proportionality between $M_D$ and
$M_U$ would not allow for the presence of a nontrivial CKM matrix
for flavor or CP violation. We postulate that the observed
$u$-quark mass patterns, along with the mismatch between the $u$-
and the $d$-quark bases, are induced by additional effective
operators of dimension $\ge 5$. For this to occur, it is
sufficient to assume the presence of a new kind of bi-doublet with
non-zero $B-L$, denoted by $\rho \sim (2,~2,~2)$, namely
\beq
\label{eq:rho} \rho = \left(\begin{array}{cc}
                \rho_1^+& \rho_2^{++} \\
                \rho_1^0 & \rho_2^{+} \\
              \end{array} \right)~,~~~
\mbox{with }\rho \stackrel{\rm Z_4}{\longrightarrow} -i \rho~.
\eeq
As $\rho$ is charged under $B-L$, it can couple to the
$\Delta_{L,R}$ fields in a $G_{LR}$- (and $Z_4$-) invariant way,
giving rise to dimension-5 operators. The Yukawa couplings for
this model, invariant under $Z_4$, can be written as:
\begin{widetext}
\beq
\label{eq:EFT5}
{\mc L}_Y &=& \mc L_{Y,Q} ~+~ \mc L_{Y,L}~, \nn \\
\mbox{with}&&\mc L_{Y,Q} ~=~ \overline{\hat Q}_L \, h \phi \, \hat Q_R~+~
\frac{1}{M_\rho} \left(
\overline{\hat Q}_L \, h_{\rho} \,\tilde\rho \Delta_R \, \hat Q_R
~+~ \overline{\hat Q}_L \, h'_{\rho} \, \Delta_L^\dagger \rho \, \hat Q_R
\right) ~+~ L \leftrightarrow R~,
% \phantom{\mbox{with}}&&\mc L_{Y,L} ~=~
% \overline{L}_L \, \tilde h_\ell \tilde\phi \, L_R
% ~+~ h'_\ell L L \Delta_L ~+~{\rm O}(1/M_\rho) ~+~ L \leftrightarrow R~,
\eeq
\end{widetext}
where all higher dimensional operators are suppressed by a new scale $M_\rho$.
Taking into account the symmetry requirements (\ref{eq:z4def}), analogous
terms can be written for the leptonic Yukawa Lagrangian, $\mc L_{Y,L}$, that
is however not relevant to the rest of our discussion.

Note that under parity
we assume that $\rho\rightarrow \tilde \rho^{\dagger}$ with all
other definitions as usual (see e.g. \cite{Zhang}), so that all the
Yukawa couplings can be made parity invariant.
Below we analyze the effects on quark
masses of adding these higher dimensional operators. In
particular,  since $v_L$ is at most of the order of the
left handed neutrino masses, we will drop the terms involving
$\Delta_L$ in analyzing fermion masses.

\subsection{\boldmath Vacuum expectation value of the $\rho$ field}\label{sec:VEVrho}

\noi A key ingredient of our discussion is that the $\rho$ field
acquires a vev and will contribute to quark masses.
However, unlike the generic situation where the mass of the
physical Higgs and its vev are of the same order, the $\rho$ vev
is induced by a tree-level tadpole, so that its mass is much
larger, in the multi-TeV range, than its vev, instead of O(100~GeV). In fact,
the discrete symmetry of our model allows a Higgs potential of the
following form:
\beq
\label{eq:VHiggs_u}
V(\phi, \Delta_R,
\Delta_L, \rho) &=& V_0(\phi, \Delta_R, \Delta_L) \nn \\
~+~ M_\rho^2
\tr(\rho^\dagger \rho) &+& (M' \tr(\phi^\dagger \tilde \rho
\Delta_R) + \hc)~,~~~~~
\eeq
where $V_0$ is the general Higgs
potential for $\phi, \Delta_R$ and $\Delta_L$ fields discussed in
ref. \cite{boris}. \footnote{\label{foot:deshpande}See Appendix of Deshpande {\em et al.}
in ref. \cite{boris}: in particular, $\mu_2, \lambda_4, \alpha_2 \to 0$
in view of our $Z_4$ symmetry, and the $\beta_i$ terms are irrelevant
to our discussion.} The mass $M^2_\rho > 0$ and minimization of the
potential (\ref{eq:VHiggs_u}) with respect to the $\rho$ vev gives
\beq
\label{eq:dPot/dvrho}
v_\rho = \frac{\ka v_R}{\sqrt2
M_\rho}\frac{M'}{M_\rho}~,
\eeq
where $\< \rho^0\> =
v_\rho/\sqrt2$ and the other vevs are defined in eq. (\ref{eq:vevs}).
As far as the value of $v_\rho$ is concerned, eq. (\ref{eq:dPot/dvrho})
shows that it depends on
the dimensional coupling $M'$ associated with the $\rho\phi\Delta_R$
term. In particular, depending on the ratio $M'/M_\rho$, the value of 
this vev can be either smaller or of order of the weak symmetry breaking 
scale.
More details can be found in section \ref{sec:numerical}.

\section{Quark mixings and FCNH Lagrangian} \label{sec:quarkmix}

\noi The Yukawa Lagrangian in eq. (\ref{eq:EFT5}) gives rise to two
contributions to quark masses. These contributions arise from the
interactions of the neutral Higgs bosons, that get a vev. Three of them
are relevant to our discussion: $\phi^0_1$, $\phi^0_2$ and $\rho^0$.
Prior to symmetry breaking, their interactions can be written as:
\beq
\label{eq:LYmass}
{\mc L}_Y &=& \overline{\hat U}_L \, h \phi^0_1 \, \hat U_R
~+~ \overline{\hat D}_L \, h \phi^0_2 \, \hat D_R \nn \\
&-& \frac{1}{M_\rho}
\overline{\hat U}_L \, h_\rho \, \rho^{0*} \delta_R^0 \, \hat U_R~,
\eeq
where, we recall, hats over the quark fields indicate flavor
eigenstates. After symmetry breaking, the terms in eq. (\ref{eq:LYmass})
give rise to the following mass terms
\beq
\label{eq:Mhat}
M_{\hat U} &=& \frac{1}{\sqrt2} \left( h \kappa ~+~ \tilde{h}_\rho v_\rho \right)~,
\nn \\
M_{\hat D} &=& \frac{h \kappa'}{\sqrt2}~,
\eeq
where we introduced the coupling
\beq
\label{eq:tildeh}
\tilde{h}_\rho \equiv - \frac{h_\rho v_R}{\sqrt2 M_\rho}~.
\eeq
Being in the flavor eigenbasis, the mass matrices $M_{\hat U, \hat D}$
are still off-diagonal.
We will denote the mass eigenbasis without the hat, and the diagonal
mass matrices as $M_{U,D}$.

Without loss of generality, we can choose a basis where $h$ is
diagonal and hence so is the down quark mass matrix -- so that
$\hat D~=~D$. There is no further diagonalization of the down quark
states. Therefore, the associated neutral Higgs boson $\phi^0_2$
has only diagonal coupling to down quarks and as such does not
lead to any flavor changing effects.

The CKM angles -- and the corresponding flavor changing
interactions -- arise then from the couplings $h$ and
$\tilde{h}_\rho$ when the combination $h\kappa \, + \,
\tilde{h}_\rho v_\rho$ is diagonalized to go to the mass
eigenbasis for the up quarks. To study these interactions, we will
henceforth neglect terms where the $\delta_R^0$ field is
dynamical, and only keep the Yukawa couplings proportional
to its vev. The first step is to redefine the two neutral-Higgs
components acting in the up-quark sector:
\newcommand{\kar}{\kappa_\rho}
\beq
\label{eq:H0basis}
H^0_1 &=& \frac{1}{\kar}(\kappa \phi^0_1 + v_\rho \rho^{0*})~,
\\ \nn
H^0_2 &=& \frac{1}{\kar}(v_\rho \phi^0_1 - \kappa \rho^{0*})~,
~~~\mbox{with}~~\kar = \sqrt{\ka^2 + v_\rho^2}~,
\eeq
such that $H^0_1$ and $H^0_2$ are orthogonal to each other, and
$\<H^0_2\>=0$. The advantage of this basis is that, when we
diagonalize the up-quark mass matrix, all the FCNH effects reside
in the $H^0_2$ coupling, whereas the $H^0_1$ coupling becomes
diagonal and does not contribute to FCNH effects.

In this basis, the neutral Higgs couplings for the up-quark mass
eigenstates read
\begin{widetext}
\beq
\label{eq:LYU}
{\mc L}_{Y,U} ~=~ \frac{\sqrt2}{\kar}
\overline{U}_L \left[ M_U \left( H_1^0 - \frac{\ka}{v_\rho} H_2^0
\right) \right]U_R ~+~ \frac{\sqrt2 \kar}{\kp v_\rho} \overline
U_L \left( V^{\rm CKM}_L M_D V^{\rm CKM \dagger}_R H_2^0 \right)
U_R ~+~ \hc~,
\eeq
\end{widetext}
with $V^{\rm CKM}_R = V^{\rm CKM}_L$ because we are in a `manifest'
LR symmetric scenario, where CP is not violated spontaneously \cite{Zhang}.
By substituting the $H^0_{1,2}$ vevs, one
immediately sees that only the first term on the r.h.s.
contributes to quark masses. Furthermore, only the second term on
the r.h.s. is flavor off-diagonal. In the following section, we
use this Lagrangian to discuss the FCNH effects and their
decoupling from the $W_R$ mass scale.

\section{FCNH effects} \label{sec:FCNH}

\subsection{Flavor changing Higgs admixture}

\noi From eq. (\ref{eq:LYU}) it is clear that FCNH effects decouple with the
mass of $H_2^0$, which is an admixture of $\phi^0_1$ and $\rho^0$. We should
therefore first discuss the neutral-Higgs mass matrix in the rotated basis.

Defining $\Phi = \{ \phi^0_1, \rho^{0*} \}$ as having square-mass matrix
$\Phi^\dagger M_\Phi^2 \Phi$, from the Higgs potential in eq. (\ref{eq:VHiggs_u}) one
finds
\beq
M_\Phi^2 =
\left( \begin{array}{cc}
~2 \lambda_1 \ka^2 + \frac{v_R^2}{2}\frac{M'^2}{M_\rho^2}~ & ~-\frac{M' v_R}{\sqrt{2}}~ \\
-\frac{M' v_R}{\sqrt{2}} & M_\rho^2 \\
\end{array}\right)~.
\eeq
While in general $\phi^0_{1,2}$ mix as well, in the limit of $\kappa
\gg \kappa'$ relevant here (see sec. \ref{sec:numerical}),
we can ignore those mixing effects and write the
above matrix by itself. The `flavor' basis in eq.
(\ref{eq:H0basis}), $H = \{ H_1^0, H_2^0 \}$, is obtained from the
$\Phi$ basis through the rotation
\beq
R \Phi = H~,~~~\mbox{with}~R = \frac{1}{\kar}\left(\begin{array}{cc}
\ka & v_\rho \\
v_\rho & - \ka \\
\end{array}\right)~,
\eeq
hence the $H$ basis has square-mass matrix $M_H^2 = R M_\Phi^2 R^T$. In particular,
the mass of the $H_2^0$ -- the particle mediating FCNH effects -- reads
\beq
\label{eq:MH20}
M_{H_2^0}^2 ~=~ M_\rho^2 + \frac{v_R^2}{r_M^2}
\left( 1 + \lambda_1 \frac{\ka^2}{M_\rho^2} + \frac{1}{4 r_M^2} \frac{v_R^2}{M_\rho^2}\right)~
\eeq
where we have defined $r_M = M_\rho / M'$, and $\lambda_1$ is one of the parameters of the Higgs
potential (see footnote \ref{foot:deshpande}).

\subsection{\boldmath Data: $D^0 - \overline D^0$ mixing}

\noi Up-quark FCNH effects can be bounded through $D^0 - \overline
D^0$ mixing. The coupling relevant to this process is the one
appearing in the $U$-$U$-$H_2^0$ interaction in eq.
(\ref{eq:LYU}), namely
\beq
\label{eq:hUUH2}
h_{U U H_2^0} ~=~
\frac{\sqrt2 \kar}{\kp v_\rho} V^{\rm CKM}_L M_D V^{\rm CKM
\dagger}_R~.
\eeq
This coupling contributes to $D^0 - \overline D^0$ mixing via tree-level
diagrams with exchange of $H_2^0$. A naive, order-of-magnitude, evaluation 
of these diagrams can be made by using the Goldstone theorem and the
vacuum saturation approximation. The result is \footnote{\label{foot:naive}%
The factor of 1/4 in eq. (\ref{eq:DeltaMDbound}) comes simply from the 
normalization of the helicity projectors as $(1 \pm \gamma_5)/2$ in the 
LR$\times$RL operator, which in turn enters twice. (Note instead that
the contributions from the other helicity combinations cancel \cite{LRLR-cancel}.)
Needless to say, this calculation omits a number of effects that in 
general are important, in particular the RGE enhancement of pseudoscalar operators.}
\beq
\label{eq:DeltaMDbound}
\Delta M_D ~\approx~ 2 \times
\frac{(h_{UU H_2^0})_{12}^2}{M_{H_2^0}^2} \cdot \frac{1}{4}
\frac{m_D^3 f_D^2}{(m_u +m_c)^2} \nn \\
~\lesssim~ \Delta M_D^{\exp}~,
\eeq
where we take $\Delta M_D^{\exp} ~=~ x / \tau ~\simeq~
1.6 \cdot 10^{-14}~{\rm GeV}$, using $x \approx 0.97 \cdot
10^{-2}$ \cite{HFAG08,AsnerPDG} and $\tau = 410 \cdot 10^{-15}~s$
\cite{PDG}. Note that in eq. (\ref{eq:DeltaMDbound}) we are
assuming the experimental result for $x$ to be saturated by
new-physics contributions. This approach is justified, given the
very poor knowledge of the SM contribution to $\Delta M_D$
\cite{D0mix_SM}. Our results would anyway barely change if we assumed, e.g.,
$\Delta M_D \lesssim 0.5 \cdot \Delta M_D^{\rm exp}$.

Plugging eq. (\ref{eq:hUUH2}) into eq. (\ref{eq:DeltaMDbound}),
and recalling that
\beq
\label{eq:assump}
&&(V^{\rm CKM}_L M_D V^{\rm CKM \dagger}_R)_{12} \simeq
m_s \lambda = 0.022~{\rm GeV}~,\nn \\
&&\frac{1}{2} \frac{m_D^3 f_D^2}{(m_u +m_c)^2} \simeq
0.1~{\rm GeV}^3~,
\eeq
the naive bound in eq. (\ref{eq:DeltaMDbound}) can be rewritten as
\beq
\label{eq:ubound}
&&v_R ~\gtrsim~ \frac{2 m_s \lambda}{B_U} \frac{\kar}{\ka \kp}
\frac{M_\rho}{M_{H_2^0}} \frac{M_\rho}{M'}~,\nn \\
&&\mbox{with}~B_U =
\sqrt{\frac{1.6 \cdot 10^{-14}~{\rm GeV}}{0.1~{\rm GeV}^3}}~.
\eeq
In the next section, we will discuss a refined version of this bound
in the context of our numerical analysis.
However, the bound in eq. (\ref{eq:ubound})
turns out to be very accurate. To get a numerical idea of this bound,
taking $\ka/\kp \approx 35$ and $\kar \approx \ka$, one would obtain
\beq
\label{eq:ubound_example}
v_R ~\gtrsim~ (15~{\rm TeV}) \, \frac{M_\rho}{M_{H_2^0}} \frac{M_\rho}{M'}~.
\eeq
The crucial point here is that the mass ratios on the r.h.s. of eq.
(\ref{eq:ubound}) can very easily provide a suppression factor of O(10)
or even larger, so that $v_R$, and hence $M_{W_R} = g_R v_R$, is in the
ballpark of 1 TeV.

\subsection{Numerical Analysis} \label{sec:numerical}

\noi In order to test quantitatively the mechanism described above, we will
now carry out a numerical exploration of the allowed parameter space of the model.
To this end, the discussion in the previous sections allows to identify
the following requirements:
\begin{enumerate}

\item The possibility we are mostly interested in, is that the scale of LR-symmetry breaking, $v_R$, be
within LHC reach. Therefore we will enforce \cite{gninenko}
\newcommand{\TeV}{~{\rm TeV}}
\newcommand{\GeV}{~{\rm GeV}}
\beq
\label{eq:MWR}
M_{W_R} = g_R v_R \lesssim 4 \TeV~,
\eeq
with $g_R = g_L$ identified with the SM SU(2)$_L$ coupling, implying $g_R \simeq 0.65$.

\item Barring accidental cancellations, which are unlikely in view of the hierarchical structure of
the CKM matrix, one expects $(M_{\hat U})_{33} \, / \, (M_{\hat D})_{33}$ (see eq. (\ref{eq:Mhat}))
to be of order $m_t/m_b \approx 40$. Therefore, we shall require
\beq
\label{eq:mt/mb}
&&\Delta_{tb} ~\equiv~ \nn \\
&&\frac{\ka}{\kp} \left( 1 \pm \left| \frac{(h_\rho)_{33}}{(h)_{33}} \right|
\left( \frac{v_R}{\sqrt2 M_\rho} \right)^2 \frac{M'}{M_\rho} \right) \nn \\
&&\hspace{20ex}~=~ {\rm O}(m_t/m_b)~.
\eeq
Taking into account the many uncertainties, we consider the range $\Delta_{tb} \in [ 27, 60 ]$ a
reasonable choice.

\item Our discussion is based on the quark--quark--neutral-Higgs Lagrangian in eq. (\ref{eq:LYmass}).
In order to justify the negligibility of contributions from operators of dimension higher than 5
allowed by the symmetry (\ref{eq:z4def}), we should require that the ratio between the dimension-5
and the dimension-4 contributions to $M_{\hat U}$ (see eq. (\ref{eq:Mhat})) be small enough.
Specifically, we will take
\beq
\label{eq:dim5/dim4}
&&\Delta_{5} ~\equiv~ \nn \\
&&\frac{(\tilde h_\rho)_{33} v_\rho}{(h)_{33} \ka} ~=~
\frac{(h_\rho)_{33}}{(h)_{33}} \frac{M'}{M_\rho} \left( \frac{v_R}{\sqrt2 M_\rho} \right)^2
\nn \\
&&\hspace{27ex}~\lesssim~~ 0.5~.
\eeq
Values of $\Delta_{5}$ even sensibly below this bound will turn out to be very easy to achieve,
as shown in the plots to follow.

\item Finally, we shall enforce the FCNH bound from $D^0 - \overline D^0$ mixing. We have implemented
a detailed FCNH calculation of $\Delta M_D$, including running effects from $M_{H_2^0}$ to $m_c$, etc.,
and using inputs from refs. \cite{D0mix_NP}. The final result reads \footnote{%
A straightforward way to obtain this result consists in implementing the (very convenient) formulae 
reported in Golowich {\em et al.}, ref. \cite{D0mix_NP}, in particular their eqs. (14), (82) and (83), 
with the strong coupling $\alpha_s$ evaluated at the different scales using {\tt RunDec} \cite{RunDec}.
Concerning $\alpha_s$ at the heavy-Higgs threshold, we chose $\alpha_s(10 \TeV)$.
}
\beq
\label{eq:accurateDeltaMDbound}
\Delta m_D ~=~ \frac{(h_{UU H_2^0})_{12}^2}{M_{H_2^0}^2} \times 0.502 \GeV^3~,
\eeq
displaying a substantial enhancement with respect to the naive bound in eq. (\ref{eq:DeltaMDbound}).
As anticipated in footnote \ref{foot:naive}, this enhancement is expected from RGE running effects
between the $M_{H_2^0}$ and $m_c$ scales. In fact, we have checked that, identifying the Wilson coefficients
at the matching scale with those at the charm scale, we get back basically the same result as eq. 
(\ref{eq:DeltaMDbound}). We will use eq. (\ref{eq:accurateDeltaMDbound}) as our FCNH constraint.

\end{enumerate}
It is clear that the main model parameters are
\beq
\label{eq:modelpars}
\mc P ~\equiv~ \{ v_R, M_\rho, M', \ka \}~,
\eeq
the other vev's or vev combinations $v_\rho$, $\kar$ and $\kp$ being fixed once the above parameters are.
Note, in particular, that the identification of $W_L$ with the SM $W$-boson allows to univocally set
the $\kp$ value as follows
\beq
\label{eq:k=v}
M_W^{\rm SM} ~=~ \frac{g v}{2} ~=~ M_{W_L} ~=~
g_L \frac{\sqrt{\ka^2 + \kp^2}}{2}~,
\eeq
where $v \simeq 250$ GeV. Note as well that the constraints (\ref{eq:mt/mb}) and (\ref{eq:dim5/dim4})
together imply $\ka$ very close to $v$ and $\kp$ much smaller, of order $\ka / \Delta_{tb}$.

The mass scale of the flavor changing neutral-Higgs admixture is in principle a further free parameter,
given that the coupling $\lambda_1$ (see eq. ({\ref{eq:MH20}})) is free. However, we shall take
$\lambda_1 \in [ 0.1, 20 ]$. With this requirement, the $M_{H_2^0}$ mass is again fixed once the
set of parameters $\mc P$ is.

Finally, the constraints in eqs. (\ref{eq:mt/mb}) and (\ref{eq:dim5/dim4}) depend on the couplings ratio
$(h_\rho)_{33}/(h)_{33}$. In order to absorb any suppression/enhancement effect into the dimensionful
parameters of the model, we shall demand this couplings ratio to be broadly of order one. Conservatively
we will require $(h_\rho)_{33}/(h)_{33} \in [ 0.1, 10 ]$.

\bigskip

\begin{figure*}[t]
\begin{center}
\includegraphics[width=0.38\textwidth]{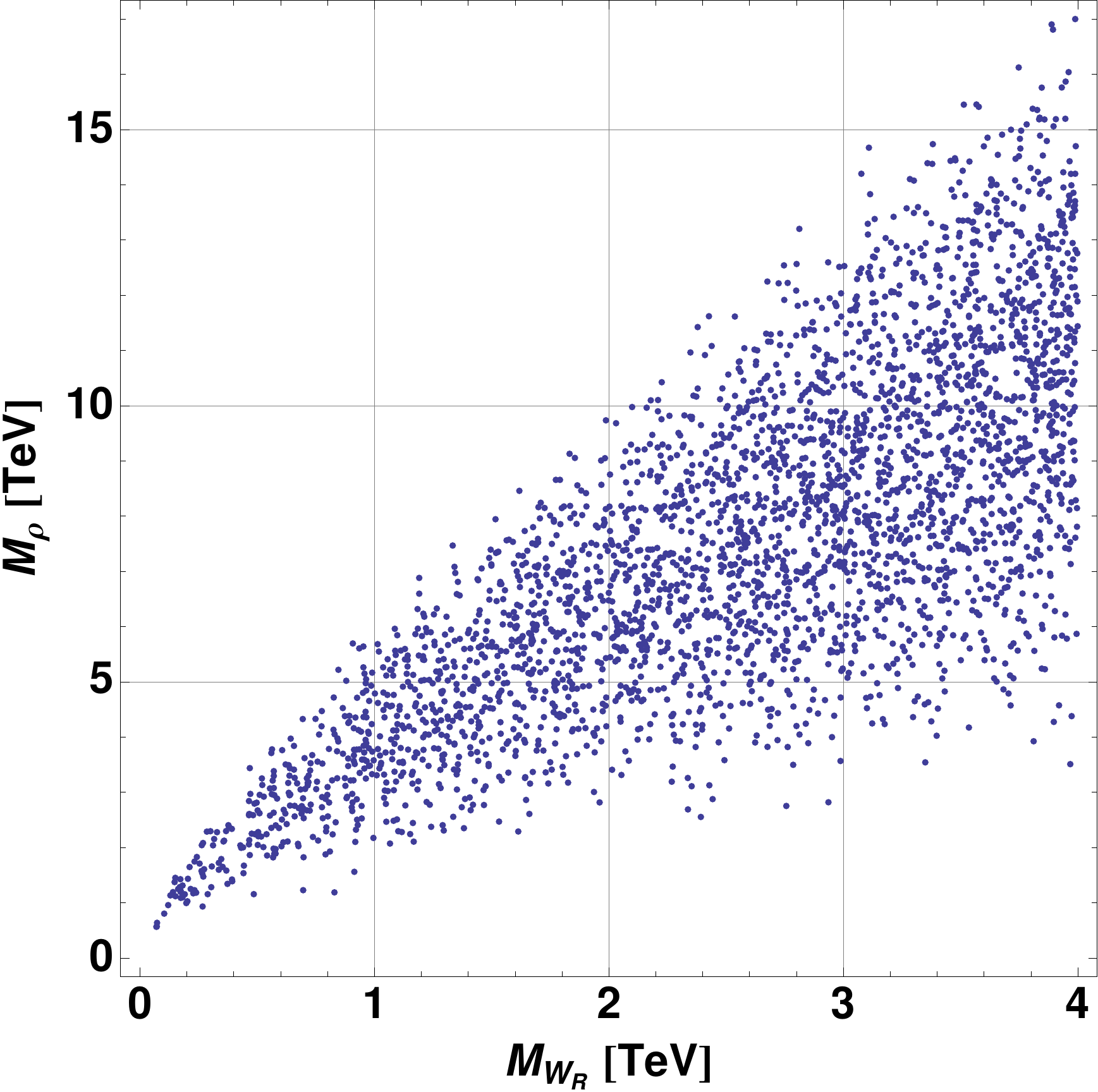} \hspace{1cm}
\includegraphics[width=0.38\textwidth]{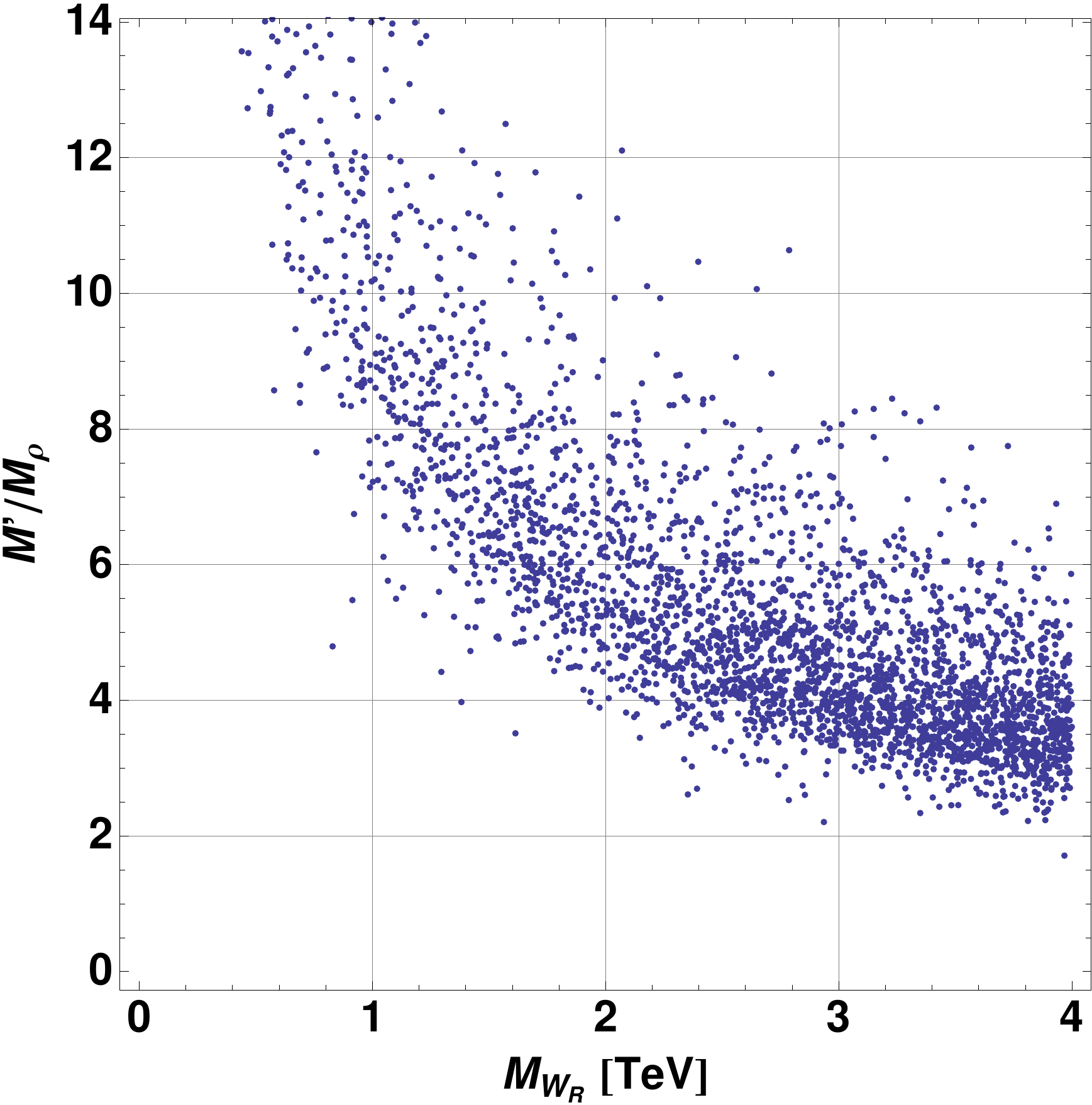}

\vspace{0.3cm}
\includegraphics[width=0.38\textwidth]{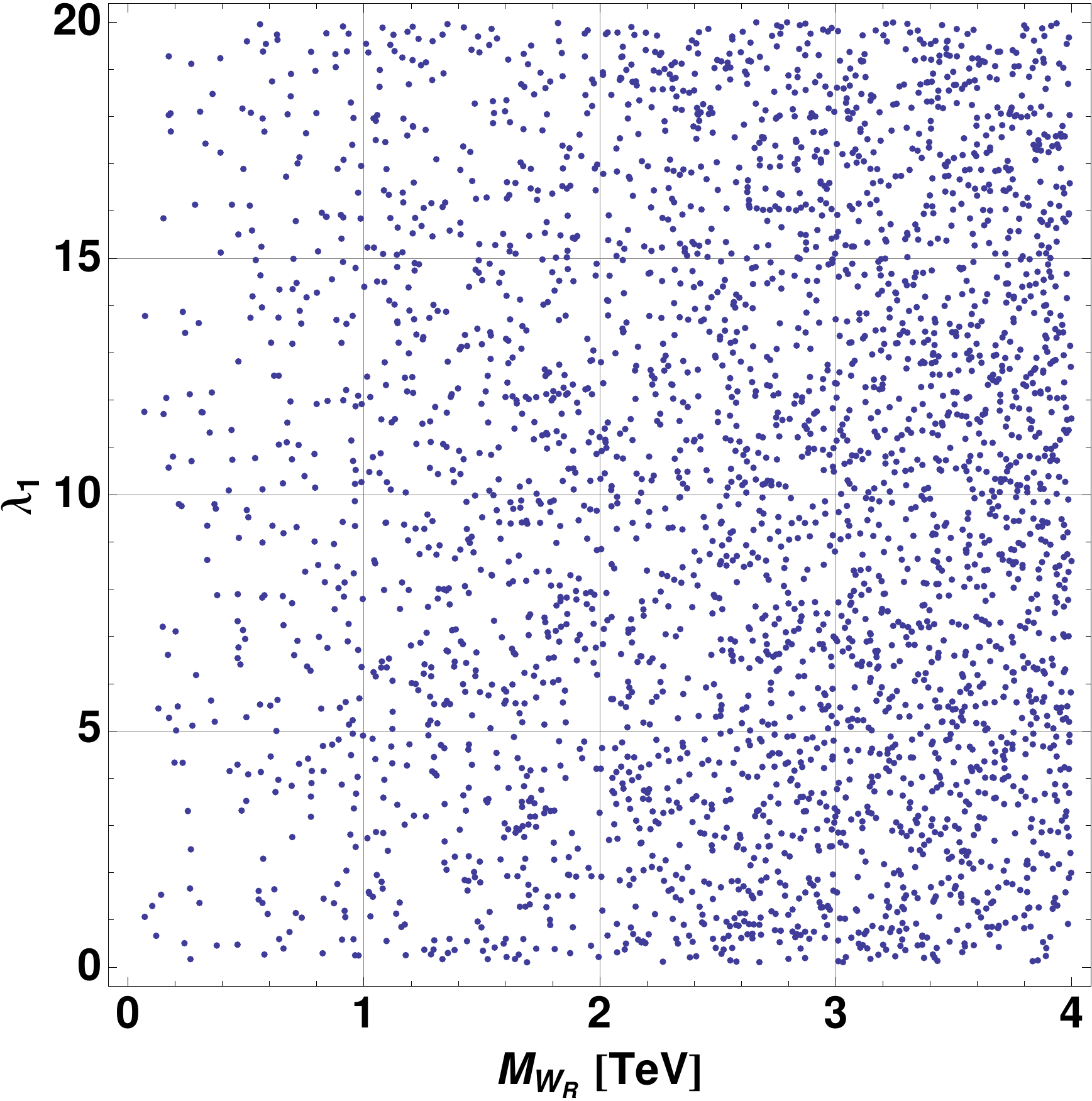} \hspace{1cm}
\includegraphics[width=0.38\textwidth]{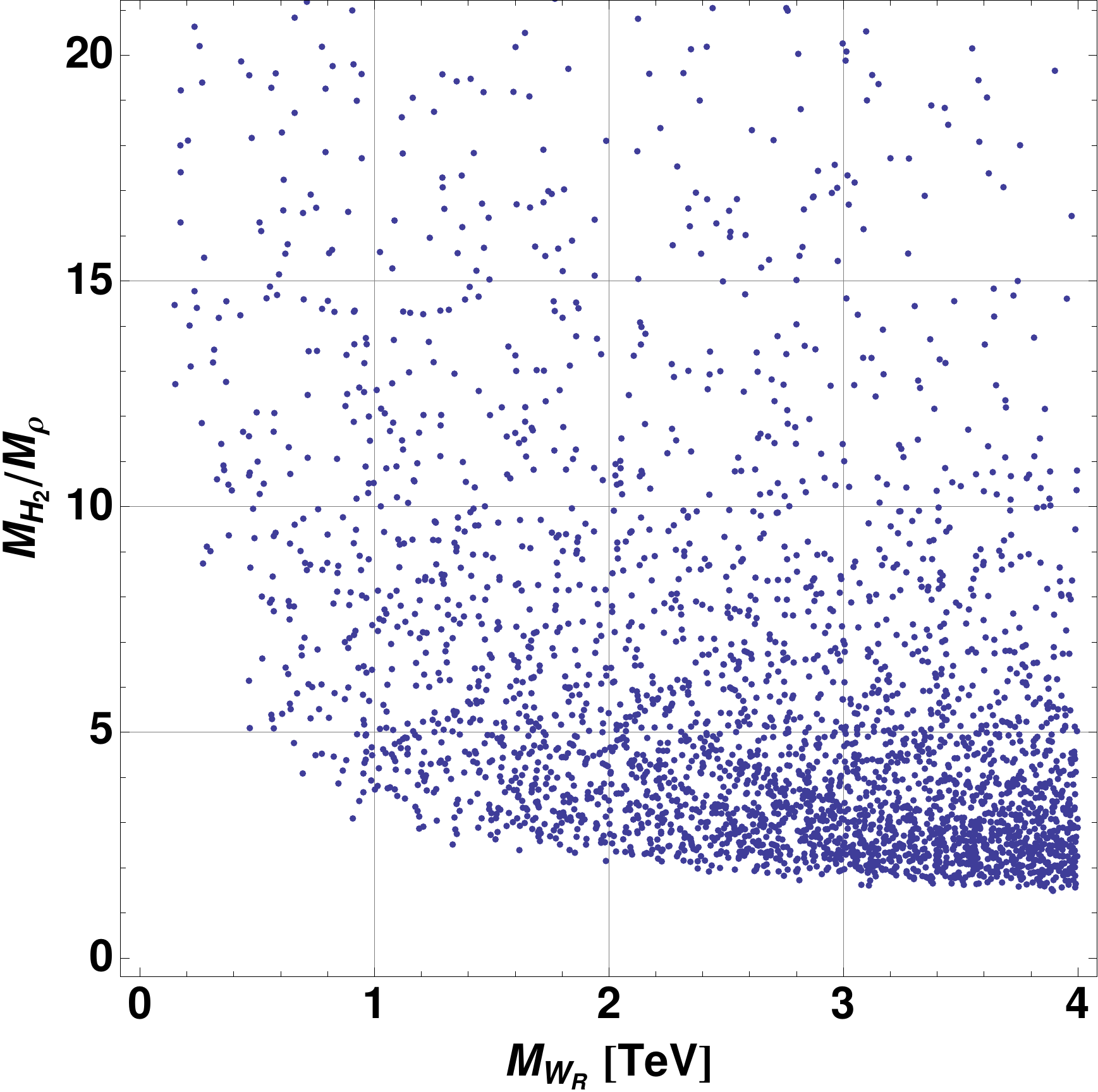}

\vspace{0.3cm}
\includegraphics[width=0.38\textwidth]{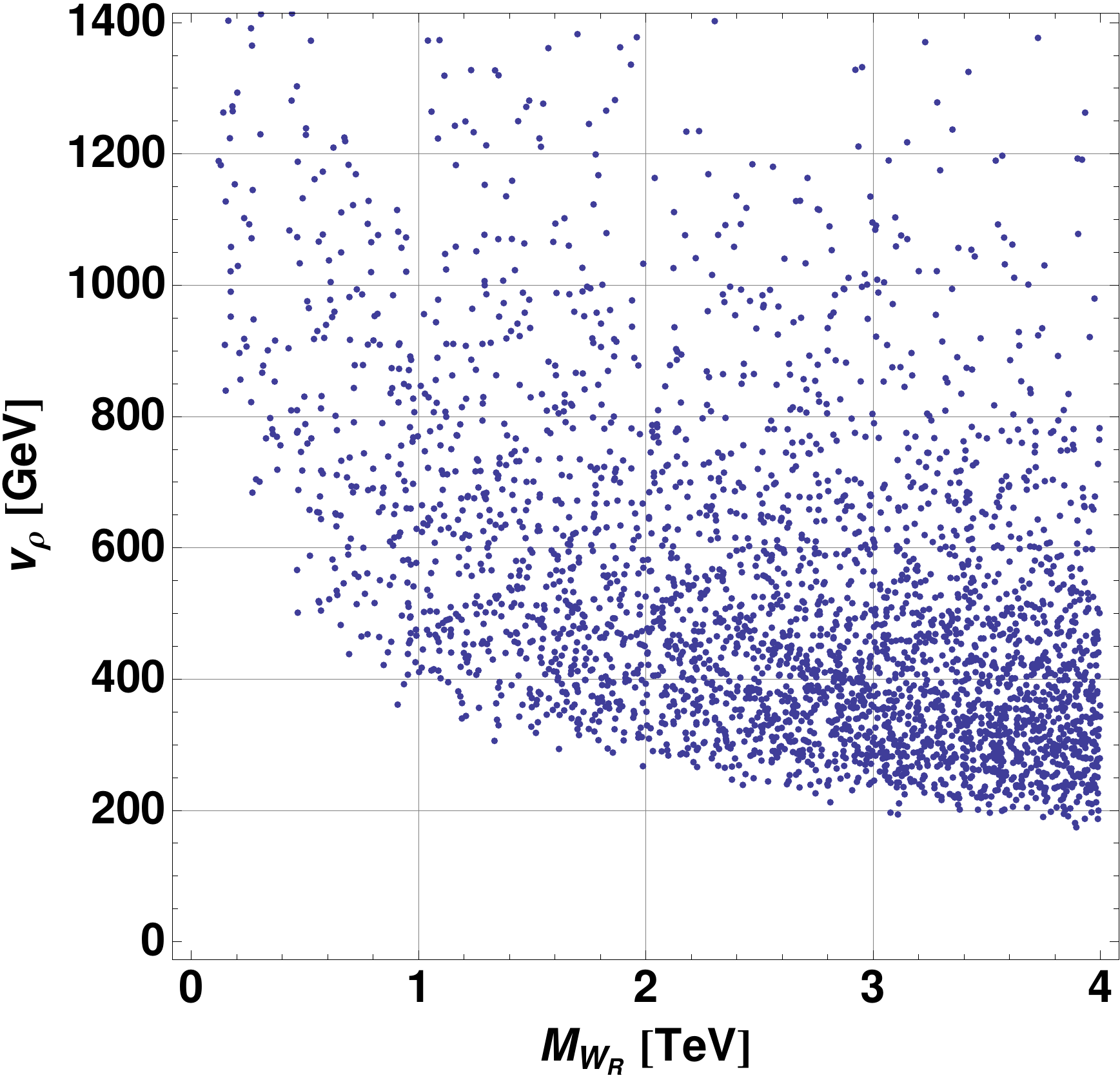} \hspace{1cm}
\includegraphics[width=0.38\textwidth]{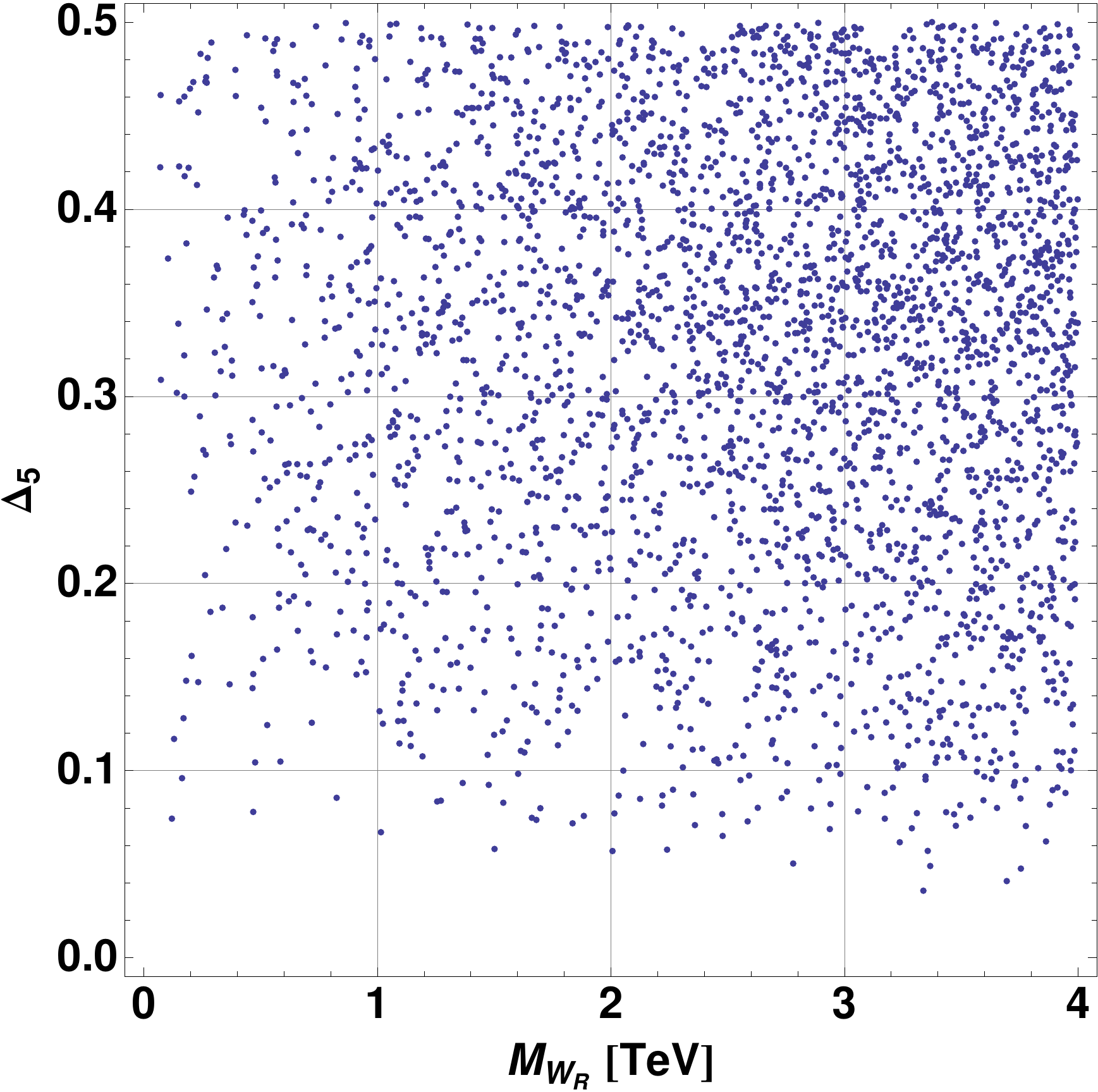}

\end{center}
\caption{Allowed parameter space for our model. See text, sec. \ref{sec:numerical}, for comments on these plots.}
\label{fig:scan}
\end{figure*}

The results of our numerical analysis are shown in fig. \ref{fig:scan}. The six panels display the $M_{W_R}$
scale as a function of all the relevant parameters discussed above, namely $M_\rho$, $M'/M_\rho$, $\lambda_1$,
$M_{H_2^0}$, $v_\rho$ and the quantity $\Delta_5$ defined in eq. (\ref{eq:dim5/dim4}).

Fig. \ref{fig:scan} shows in the first place that $M_{W_R}$ values in the range
1 to 4 TeV -- well within the LHC reach \cite{gninenko} -- are very natural to
obtain. This requires $M_\rho$ in the O(10 TeV) range, and values for $M'$
and for $M_{H_2^0}$ higher than about $2 M_\rho$, but not hierarchically higher. The fact that no fine
tuning in these masses is needed is confirmed by the values for $\lambda_1$ -- which, according
to eq. (\ref{eq:MH20}), can be traded for $M_{H_2^0}$. The allowed $\lambda_1$ values
in fig. \ref{fig:scan} are quite uniformly distributed in their whole allowed range,
for $M_{W_R}$ in basically the entire interval $[1,4]$ TeV.

In addition, the $M_{W_R}$ vs. $v_\rho$ panel shows (as a consistency check not imposed in the scan)
that $v_\rho < v_R$ is always fulfilled, justifying in turn our approximation of neglecting FCNH effects due
to the propagation of $\delta_R^0$ with respect to those due to $\rho^0$.

Finally, as a further consistency check, the $M_{W_R}$ vs. $\Delta_5$ panel shows that the ratio $\Delta_5$
(see definition in eq. (\ref{eq:dim5/dim4})) between dimension-5 and dimension-4 contributions to the
top mass is very naturally well below 1.

\section{Discussion}\label{sec:comments}

\noi Here we collect some further remarks on the model.

\begin{itemize}

\item[{\em (i)}] In the leptonic sector, not discussed in this
Letter, the renormalizable $\rho$ couplings must be chosen fairly
close to diagonal form, because all the neutrino mixings must
arise from the right handed neutrino mass matrix to be consistent
with lepton flavor violating branching ratios such as $\mu\to 3e$
etc.

\item[{\em (ii)}] Our idea can be extended to the supersymmetric
left-right models and is particularly appealing in this case. The
point is that, due to holomorphicity of the superpotential,  in
SUSYLR models one needs automatically two bi-doublets (in place of
$\phi$ alone) to generate nontrivial CKM angles. Using our idea,
one can now replace the second $B-L=0$ bi-doublet by a $B-L=2$
bi-doublet and use a higher dimensional operator of the form
$\frac{h_\rho}{M}Q^T\rho\Delta^cQ$ to generate CKM angles as well
as to solve the FCNH problem while keeping the $W_R$ scale in the
LHC accessible range. No cancellations between Higgs exchange
contribution and squark box graphs need be invoked \cite{other1}.

\item[{\em (iii)}] The new $\rho$ particles of the model are much
heavier than the parity breaking scale and are therefore beyond
the reach of the LHC.

\item[{\em (iv)}] The $M'$ scale in the potential in eq.
(\ref{eq:VHiggs_u}) could arise from the vev of a field $\sigma$
reflecting higher scale physics, all our results remaining
unchanged.

\end{itemize}

\section{Conclusion}

\noi In summary, we have presented an extension of the left-right
symmetric model which satisfies the bounds on Higgs masses arising from
flavor changing effects, without at the same time dragging the parity
breaking scale up with it. Within our model,
the contributions to quark masses that break the relation 
${\rm diag}(m_u,m_c,m_t)$ $\propto$ ${\rm diag}(m_d,m_s,m_b)$,
as well as flavor mixing ($V^{\rm CKM} \neq \id$), are both the effect
of a dimension 5 operator. This keeps the $W_R$ and $Z'$ within
the reach of LHC and makes it possible, as the LHC collects
data, to explore both the origin of parity violation and of neutrino
masses -- a common origin, in the context of our class of models.
The added new particles are, on the other hand, beyond the LHC reach.

\subsection*{Acknowledgements}

\noi The work of R. N. M. is supported
by the U. S. National Science Foundation grant No. PHY-0652363.
R. N. M. would like to acknowledge the hospitality of Michael Ratz,
and of Technical University of Munich and Excellence Cluster Universe,
where this work was started. The work of DG is supported by the
DFG Cluster of Excellence `Origin and Structure of the Universe'.

\end{document}